\documentstyle[12pt]{article}
\title{Excitation of nuclear anharmonic vibrations in heavy-ion 
fusion reactions \\}
\author{K. Hagino$^{1}$, S. Kuyucak$^{2}$, and N. Takigawa$^{1}$ \\ \\
\medskip
{\it $^{1}$Department of Physics,
Tohoku University, Sendai 980--77, Japan}
\\
{\it $^{2}$Department of Theoretical Physics, Research School of Physical
Sciences, } \\
{\it Australian National University, Canberra,
ACT 0200, Australia}
}
\date{}

\begin{document}

\maketitle

\begin{center}
{\bf Abstract}
\end{center}

We discuss the effects of multi-phonon excitations on heavy-ion fusion 
reactions at energies near and below the Coulomb barrier, focusing 
especially on the role of anharmonicities.  We 
carry out a systematic study of those effects on the excitation function 
of the fusion cross section and on the fusion barrier distribution, by 
using the vibrational limit of the interacting boson model.  We 
also analyze the recently measured high-precision data of the $^{16}$O + 
$^{148}$Sm fusion reaction with this model and discuss the anharmonic 
properties of the quadrupole as well as the octupole vibrations in 
$^{148}$Sm.  Negative and positive static quadrupole moments are deduced 
for the first 2$^+$ and 3$^-$ states in $^{148}$Sm, respectively.  It is 
shown that the fusion barrier distribution 
strongly depends on the sign of the quadrupole 
moments, suggesting that subbarrier fusion reactions offer an 
alternative method to extract the static quadrupole moments of phonon 
states in spherical nuclei.

\medskip

\noindent
PACS number(s): 25.70.Jj, 21.60.Fw, 21.10.Ky, 27.60.+j

\newpage

\begin{center}
{\bf I. INTRODUCTION}
\end{center}

Large enhancement of heavy-ion fusion cross sections at 
subbarrier energies relative to 
predictions of a one-dimensional potential model 
has been well 
recognized \cite{B88}.  It is by now well established that these 
enhancements can be attributed to couplings 
of the relative motion between the colliding nuclei 
to their intrinsic excitations 
as well as transfer reactions \cite{BT97}.  When the excitation energy of an 
intrinsic motion is neglected, 
the corresponding channel-coupling gives rise to a 
distribution of potential barriers \cite{E81,NBT86,HTBB95}.  
Based on this idea, a method was proposed to 
extract barrier distributions directly from fusion excitation functions 
by taking the second derivative of the product of the fusion cross 
section and the center of mass energy $E \sigma$ with respect to
$E$ \cite{RSS91}.  
It was recently shown that the 
concept of the barrier distribution can be extended also to 
the cases where the intrinsic motion carries a finite excitation 
energy \cite{HTB97}.  
Coupled-channels calculations have shown 
that the fusion barrier distribution, i.e.  $d^2(E\sigma)/dE^2$, is very 
sensitive to the details of the channel couplings, while the fusion 
excitation function itself is rather featureless \cite{DHLH97}.  In 
order to deduce meaningful experimental barrier distributions and to 
observe their sensitivity to the couplings, 
the excitation function of fusion cross sections 
has to be measured with high precision at  small 
energy intervals.  Thanks to the recent developments in experimental 
techniques \cite{WLW91}, such data are now available for several 
systems, and they have clearly demonstrated that the barrier 
distribution is indeed a sensitive quantity to the channel couplings 
\cite{LDH95}.  They clearly show the effects of couplings to static 
deformations and associated rotational motions 
\cite{LDH95,WLH91,LLW93,LRL93,BCL96}, vibrational degrees of freedom 
\cite{LDH95,BCL96,MDH94,TCS97,SACH95,SACN95}, and transfer 
reactions\cite{LDH95,BCL96,MDH94,TCS97,SACH95}, in a way much more 
apparent than in the fusion excitation function itself.  These high 
precision data have thus enabled a detailed study of the effects of 
nuclear collective excitations on fusion reactions, and have generated 
a renewed interest in heavy-ion subbarrier fusion reactions.

Nuclear surface oscillations with various multipolarities are typical 
examples where the barrier distribution analyses 
of fusion reactions 
may be applied to study their detailed properties.  For instance, the 
barrier distribution analysis of the recently measured accurate data on 
the $^{58}$Ni + $^{60}$Ni fusion reaction has shown 
evidence for 
couplings to double-phonon states in $^{58}$Ni and $^{60}$Ni 
\cite{SACN95}.  The barrier distributions were shown to be quite 
sensitive to the number of phonons excited during fusion reactions.  To 
the zeroth approximation, these surface vibrations can be described in 
terms of harmonic oscillations, which 
dictate simple relations among 
the level energies and the electromagnetic transitions between them.  
Some of the characteristic features of harmonic 
oscillations are 
summarized as follows: (i) all the levels in a phonon multiplet are 
degenerate and the energy spacing between neighboring multiplets is a 
constant, (ii) the electric transitions between neighboring multiplets 
are linearly related, e.g.  B(E2; I$_1^+ \to$ (I$-$2)$_1^+$) = (I/2) B(E2; 
2$_1^+ \to $ 0$_1^+$), (iii) the static quadrupole moment is zero for 
all the phonon states.  In realistic nuclei, however, there are residual 
interactions between phonons, which cause deviations from the harmonic 
oscillator limit \cite{BTK65}.  
The anharmonicity leads to a level splitting within a multiplet and 
to modifications in the ratios between various electromagnetic 
transition strengths.  
Furthermore, the anharmonicities 
generate a finite value of static quadrupole moment in excited states 
\cite{TU66}.

In even-even nuclei near closed shells, there are many examples of 
two-phonon triplets ($0^+,2^+,4^+$) of quadrupole surface vibrations 
\cite{BM75}.  
Though the 
central position of their excitation energies 
is approximately twice the energy of the first $2^+$ state, they usually 
exhibit appreciable splitting within the triplet.  A theoretical 
analysis of the anharmonicities for the quadrupole vibrations was first 
performed by Brink et al.  \cite{BTK65}, who related the excitation 
energies of three-phonon quintuplet ($0^+,2^+,3^+,4^+,6^+)$ to those of 
the double-phonon triplets, and also gave relations between the 
electric transition strengths from the three- to the two-phonon states 
and those from the two- to the one-phonon states.  These relations, 
however, had not been confirmed until recently because of the sparse 
experimental data on three-phonon states.  The experimental situation 
has improved rapidly in recent years \cite{ABC87}, and data on 
multi-phonon states are now available for several nuclei.  As a 
consequence, study of multi-phonon states, and especially their 
anharmonic properties, is attracting much interest 
\cite{CZB93,ZC94,KGJL95,KJ95,KOG96}.  It is 
worth mentioning that 
anharmonic effects are not restricted to low-lying vibrations but 
play significant roles also 
in multi-phonon excitations of giant resonances in 
heavy-ion collisions \cite{VCC95,LAC97,BF97}.

In many even-even nuclei near closed shells, 
the first 3$^{-}$ state appears 
at a relatively low excitation energy, which 
competes with the quadrupole state \cite{BM75,MJC95}.  
These low-lying 3$^{-}$ states 
have been frequently interpreted as collective 
octupole vibrations arising from a coherent sum of one-particle one-hole 
excitations between single particle orbitals differing by three units of 
orbital angular momentum \cite{BM75,CB87}.  This picture is supported by 
the large E3 transition probabilities from the 
first 3$^{-}$ state to the 
ground state, and suggests the possibility of multi-octupole-phonon 
excitations.  In contrast to the quadrupole vibrations, however, so far 
there are only little experimental 
evidence for double-octupole-phonon 
states.  One reason for this is that E3 transitions from two-phonon 
states to a single-phonon state compete against lower multipolarity E1 
transitions.  This makes it difficult to unambiguously identify the 
two-phonon quartet states ($0^+,2^+,4^+,6^+$).  
Consequently despite the 
fact that the first 3$^-$ state of $^{208}$Pb has a large quadrupole 
moment, which indicates a significant anharmonic effect in octupole 
vibrations \cite{H70,S71,GRC75,JBF77}, a direct study of the anharmonic 
properties in multi-octupole phonon spectra has not been possible for a 
long time.  Only in recent years, convincing evidence have been 
reported for double-octupole-phonon states, as well as double-phonon 
states built from single octupole and quadrupole phonon states, in some 
nuclei, e.g., $^{208}$Pb \cite{YGM96}, $^{144}$Sm 
\cite{GVB90,GJB93,WRZB96,MHK93}, $^{147}$Sm \cite{UBJ96}, $^{146}$Sm 
\cite{BBB95}, $^{145}$Nd \cite{UBJ96}, $^{144}$Nd \cite{BBB95}, and 
$^{148}$Gd \cite{PKB93,K92}.

In Ref.~\cite{HTK97}, we have shown that heavy-ion subbarrier fusion 
reactions provide a powerful method to study the anharmonic properties 
of quadrupole as well as octupole vibrational excitations.  We analyzed 
the recently measured high precision data for the $^{16}$O + $^{144}$Sm 
reaction and extracted negative quadrupole moments for both the first 
2$^+$ and 3$^-$ states in $^{144}$Sm.  The aim of the present paper is, 
in addition to giving the full details of that analysis, to carry out a 
systematic study of the effects of anharmonicities on subbarrier fusion, 
and to apply the same analysis to the $^{16}$O + $^{148}$Sm fusion 
reaction to study the anharmonic properties of the vibrational 
excitations in $^{148}$Sm and their effects on the fusion reaction.  
Although the importance of the anharmonic effects on proton scattering 
\cite{CPM82}, as well as the reorientation effects in phonon spectra on 
heavy-ion elastic and inelastic scattering \cite{TST90,SLN90} has been 
pointed out, no systematic studies of the effects of anharmonicities on 
subbarrier fusion and on fusion barrier distributions have been 
performed so far.  In view of the high precision data on subbarrier 
fusion reactions, which have recently become available, such studies are 
necessary in discussing the effects of vibrational excitations on 
subbarrier fusion reactions.  The paper is organized as follows.  In 
Sec.~II, we formulate coupled-channels calculations which explicitly 
take into account the anharmonic properties of vibrational excitations 
using the vibrational limit of the interacting boson model \cite{IA87}.  
We compare the formalism with that in the harmonic limit and point out 
several important features of the anharmonic vibrational excitations.  
In Sec.~III, a systematic study of the effects of anharmonicities on 
fusion reactions is presented.  
They include the effects of unequal 
spacing of levels, reorientation, and finite boson number.  In Sec.~IV, 
the formalism is applied to the $^{16}$O + $^{148}$Sm fusion reaction.  
We extract the quadrupole moments of the first 2$^+$ and 3$^-$ states in 
$^{148}$Sm from the analysis of the high quality fusion data available 
for this system.  Finally, a summary of the work is given in Sec.~V.

\begin{center}
{\bf II. EFFECTS OF PHONON EXCITATIONS ON FUSION}
\end{center}

In this section, we describe 
the basic formalism for 
treating the vibrational excitation in subbarrier fusion 
reactions. 
Let us first consider the case where the relative motion between the 
colliding nuclei couples to the quadrupole vibrations 
in the target nucleus. 
The total Hamiltonian of the system is assumed to be 
\begin{equation}
H=-\frac{\hbar^2}{2\mu}\nabla^2+V_N(r)+\frac{Z_PZ_Te^2}{r}+
H_{int}+ V_{coup}({\mbox{\boldmath $r$}},\xi),
\label{ham}
\end{equation}
where ${\mbox{\bf r}}$ is the coordinate of the relative motion 
between the projectile and the target, and $\mu$ is the reduced mass.
$V_N$ is the bare nuclear potential, $Z_P$ and $Z_T$ are the 
atomic numbers of the projectile and the target, respectively. 
$H_{int}$ describes the vibrational excitations in the target nucleus,
and $V_{coup}$, the coupling between these excitations (generically 
denoted by $\xi$) and the relative motion. 
In Sub-sections A and B, we discuss the coupling of the harmonic and 
anharmonic vibrators within the linear coupling approximation.  Although 
this approximation is too simplistic to 
apply to realistic systems \cite{BBK93,HTD97a}, it enables us to 
easily understand the essential features of the 
effects of anharmonicities.  
Extensions of the model so as to include the couplings to 
all orders and to the octupole vibrations in the target nucleus 
as well are given in Sub-sections C and D, respectively.

\begin{center}
{\bf A. Harmonic limit}
\end{center}

In the geometrical model of Bohr and Mottelson, the radius of the 
vibrating target nucleus is parameterized as 
\begin{equation}
R(\theta,\phi) = R_T \left(1+\sum_{\mu}\alpha_{2\mu}
Y_{2\mu}^*(\theta,\phi) \right),
\end{equation}
where $R_T$ is the equivalent sharp surface radius 
of the target nucleus 
and $\alpha_{2\mu}$ are the coordinates of the surface vibration.  
To the lowest order, 
the surface oscillations are approximated by a harmonic oscillator and 
the corresponding Hamiltonian is given by 
\begin{equation}
H_{int}=\hbar\omega \left(\sum_{\mu} a_{2\mu}^{\dagger}a_{2\mu}
+\frac{5}{2}\right). 
\label{hint}
\end{equation}
Here $\hbar\omega$ is the oscillator quanta and $a_{2\mu}^{\dagger}$ 
and $a_{2\mu}$ are the phonon creation and annihilation operators, 
respectively. 
The surface coordinates $\alpha_{2\mu}$ and the 
phonon creation and annihilation operators are related by 
\begin{equation}
\alpha_{2\mu}=\alpha_0 \left(a_{2\mu}^{\dagger}+(-)^{\mu}a_{2\mu} 
\right),
\end{equation}
where $\alpha_0$ is the amplitude of the zero point motion. 
It is related to the quadrupole deformation parameter by 
$\alpha_0=\beta_2/\sqrt{5}$ \cite{BM75} and can be estimated from the 
experimental transition probability using 
\begin{equation}
\alpha_0=\frac{1}{\sqrt{5}}\frac{4\pi}{3Z_TR_T^2}
\sqrt{\frac{B(E2)\uparrow}{e^2}}. 
\end{equation}
In the collective model, the coupling Hamiltonian between the relative 
motion and the quadrupole surface oscillations is 
often assumed to be 
\begin{equation}
V_{coup}({\mbox{\bf r}},\alpha_2) = f(r) \sum_{\mu}\alpha_{2\mu}
Y_{2\mu}^*(\hat{{\mbox{\bf r}}}).
\label{vcoup}
\end{equation}
This is the so called linear coupling approximation where 
only the linear term with respect to the surface coordinate is taken 
into account.  
The coupling form factor, $f(r)$, 
consists of the nuclear and the Coulomb parts and is given by 
\begin{equation}
f(r)=-R_T\frac{dV_N}{dr}+\frac{3}{5}Z_PZ_Te^2\frac{R_T^2}{r^3}. 
\end{equation}

The dimension of the resulting coupled-channels 
calculations is in general 
too large for practical purposes.  
Here, we introduce two approximations 
which considerably reduce the 
dimension of the coupled-channels calculations 
with negligible effects on its accuracy.  
The first one is the no-Coriolis approximation where 
one transforms the whole system to the rotating frame where 
the $z$ axis is along the direction of the relative motion 
${\mbox{\bf r}}$ at every instant \cite{TI86,HTB95}.  
This is achieved by 
replacing the angular momentum of the relative motion in each channel by 
the total angular momentum $J$.  
As the operator for a rotational 
coordinate transformation in the whole space then commutes with the 
centrifugal operator for the relative motion, one can make this 
transformation to the rotating frame without introducing any 
complications \cite{HTB95}.  Using $Y_{2\mu}(\hat{{\mbox{\bf r}}}=0) = 
\sqrt{5/4\pi}\delta_{\mu,0}$, the coupling Hamiltonian (\ref{vcoup}) in 
the rotating frame becomes 
\begin{equation}
V_{coup}({\mbox{\bf r}},\alpha_2) = \sqrt{\frac{5}{4\pi}} 
f(r) \alpha_{20} = 
\frac{\beta_2}{\sqrt{4\pi}}f(r)
\left(a_{20}^{\dagger}+a_{20}\right). 
\label{vcoupa}
\end{equation}
Since the projection of the total angular momentum $J$ onto the $z$-axis 
of the rotating frame, i.e.  the $K$ quantum number, is conserved in 
this approximation, the dimension of the coupled-channels 
equations is drastically reduced.

A further reduction of the coupled-channels equations can be achieved by 
introducing the $n$-phonon channels \cite{TI86,KRNR93}.  For the 
quadrupole surface vibrations, the two phonon state has three levels 
($0^+,2^+,4^+$).  In the harmonic limit, this two-phonon triplet is 
degenerate in the excitation energy.  One can then replace the couplings 
to all the members of the two-phonon triplet by the coupling to a single 
state given by 
\begin{equation}
|2> = \sum_{I=0,2,4} <2020|I0> |I0> = 
\frac{1}{\sqrt{2!}}(a^{\dagger}_{20})^2 |0>. 
\end{equation}
In the same way, one can introduce the $n$-phonon channel as 
\begin{equation}
|n> = \frac{1}{\sqrt{n!}}(a^{\dagger}_{20})^n |0>. 
\label{nphonon}
\end{equation}
If we truncate to the two-phonon states, the corresponding coupling 
matrix is given by 
\begin{equation}
V_{coup}=
\left(\begin{array}{ccc}
0&F(r)&0\\
F(r)&\hbar\omega
&\sqrt{2}F(r)\\
0&\sqrt{2}F(r)&2\hbar\omega
\end{array}\right),
\end{equation}
where, $F(r)$ is defined as $\beta_2 f(r)/\sqrt{4\pi}$. 

\begin{center}
{\bf B. Anharmonic vibrator}
\end{center}

Higher order corrections to Eq.~(\ref{hint}) lead to anharmonicities in 
the vibrational excitation.  The effects of anharmonicity in surface 
vibrations can be described in many different ways.  Among them, the 
interacting boson model (IBM) \cite{IA87} in the vibrational (U(5)) 
limit provides a convenient calculational framework to discuss the 
anharmonic effects.  The vibrational limit of the IBM and the anharmonic 
vibrator (AHV) in the geometrical model are very similar, the only 
difference coming from the finite number of bosons in the former 
\cite{ABC87,CW88,G82}.  The eigenvalues of the intrinsic Hamiltonian 
$H_{int}$ in the U(5) limit are given by \cite{IA87} 
\begin{equation}
E_{n_dvI}=\epsilon n_d+\alpha n_d(n_d+4)+\beta v(v+3)+\gamma I(I+1),
\label{eu5}
\end{equation}
where $n_d$, $v$, and $I$ are the quantum numbers giving the number of 
$d$-bosons, the $d$-boson seniority, and the intrinsic angular momentum, 
respectively.  $\epsilon, \alpha, \beta$, and $\gamma$ are adjustable 
parameters.  The first term gives equally spaced and degenerate phonon 
spectra, while the splitting of multi-phonon multiplets due to the 
anharmonic effects are caused by the remaining terms.

A model for subbarrier fusion reactions, which uses the IBM to describe 
the effects of channel couplings, has been developed by Balantekin et 
al.~\cite{BBK93,BBK94,BBK94b}.  They assume that the coupling 
Hamiltonian has a similar form to that of the collective model
Eq.~(\ref{vcoup}).  In the linear coupling and the no-Coriolis 
approximations, it is given as 
\begin{equation}
V_{coup}(r,\xi)=\frac{\beta_2}{\sqrt{4\pi N}}f(r) Q_{20}. 
\label{vibm}
\end{equation}
Here, $N$ is the boson number and we have introduced the 
scaling of the coupling strength with $\sqrt{N}$ to ensure 
the equivalence of the IBM and the geometric model results 
in the large $N$ limit \cite{BBK93,BBK94,BBK94b}. 
$Q_{20}$ is the quadrupole operator in the IBM, which is given by 
\begin{equation}
Q_{20}=s^{\dagger}d_0 + sd_0^{\dagger} +
\chi_2(d^{\dagger}\tilde{d})^{(2)}_0,
\label{quad}
\end{equation}
where tilde is defined as $\tilde{d}_{\mu}=(-)^{\mu}d_{-\mu}$.  
Systematic studies of subbarrier fusion reactions in this model 
indicate that 
the coupling strengths used in this model are very similar to those in 
the geometrical model \cite{BBK94}.  We, therefore, assume that the 
coupling strength in Eq.~(\ref{vibm}) is the same as that in the 
harmonic limit of the geometrical model, Eq.~(\ref{vcoupa}).  This 
approximation is valid when the anharmonic effects are not so large, 
so as to allow description of vibrational mode of excitations in terms 
of the U(5) limit.

In anharmonic vibrators, phonon states within 
a given multiplet are split in 
energy.  Thus one has to treat each state separately 
as a different channel.  The 
introduction of the multi-phonon channel, as in the harmonic limit, is 
possible only when the multi-phonon states are degenerate in the 
excitation energies.  It will be shown in the next section that one can 
safely neglect the effects of this energy splitting 
on fusion cross section as well as on the fusion barrier distribution.  
Accordingly, in order to keep the 
calculations simple, we assume in the following that the multi-phonon 
multiplets are degenerate in their excitation energy unless explicitly 
mentioned.  The wave function of the $n$-phonon channel in the framework 
of the IBM then reads 
\begin{equation}
|n>=\frac{1}{\sqrt{n!(N-n)!}}(s^{\dagger})^{N-n}(d_0^{\dagger})^n
|0>,
\end{equation}
and the corresponding coupling matrix, truncated to the two-phonon 
states, is given by 
\begin{equation}
V_{coup}=
\left(\begin{array}{ccc}
0&F(r)&0\\
F(r)&\hbar\omega-{2 \over \sqrt{14N}} \chi_2 F(r)
&\sqrt{2(1-1/N)}F(r)\\
0&\sqrt{2(1-1/N)}F(r)&2\hbar\omega+\delta
-{4 \over \sqrt{14N}} \chi_2 F(r)
\end{array}\right).
\label{v2p}
\end{equation}
The parameter $\delta $ has been introduced 
to represent deviation of the energy spectrum from the 
equal spacing in the harmonic oscillator limit. 
When the $\chi_2$ parameter in 
the quadrupole operator is zero, quadrupole moments of all states 
vanish, and one obtains the harmonic limit in the large $N$ limit.  
Non-zero values of $\chi_2$ generate quadrupole moments and, together 
with finite boson number, are responsible for the anharmonicities 
in electric transitions.  The differences in the coupling scheme 
in the harmonic oscillator limit and in the case with anharmonicities 
are summarized schematically in Fig.~1.

\begin{center}
{\bf C. All order couplings}
\end{center}

It has been pointed out that 
the linear coupling approximation for the nuclear 
coupling is not adequate in heavy-ion fusion reactions 
\cite{BBK93,HTD97a}.  Higher order couplings have to be included in 
order to describe realistic systems and compare 
the theoretical calculations with the experimental data.  
On the other hand, 
non-linear terms were shown to be 
negligible for the Coulomb 
couplings \cite{HTD97a}.  Higher order corrections to the nuclear 
coupling were discussed in Ref.~\cite{HTD97a} in the case of a harmonic 
oscillator.  The coupling matrix element between $n$- and $m$- phonon 
channels in the harmonic limit is given by 
\begin{equation}
V_{nm}^{(N)}(r)
=\int^{\infty}_{-\infty}dxu^*_n(x)u_m(x)
\frac{-V_0}{1+\exp\left[\left(r-R_0-\sqrt{\frac{5}{4\pi}}
R_T x\right)/a\right]}. 
\end{equation}
Here we have adopted the no-Coriolis approximation and the 
nuclear potential is assumed to have a Woods-Saxon form. 
Note that this equation includes the bare nuclear potential $V_N$ 
in the total Hamiltonian (Eq.~(\ref{ham})). 
$u_n(x)$ is the wavefunction of the $n$-th excited state of the 
harmonic oscillator (see Eq.~(\ref{nphonon})) and is given by 
\begin{equation}
u_n(x)=\frac{1}{\sqrt{2^nn!}}\frac{1}{\sqrt[4]{2\pi\alpha_0^2}}
H_n\left(x/\sqrt{2}\alpha_0\right)e^{-x^2/4\alpha_0^2}, 
\end{equation}
where $H_n(x)$ is the Hermite polynomial. 

Extension of the anharmonic oscillator coupling in the 
linear order approximation to all orders can be 
carried out in a similar manner. Following 
Refs.~\cite{BBK93,BBK94,BBK94b}, we assume that the nuclear coupling 
Hamiltonian is given in the no-Coriolis approximation by 
\begin{equation}
V^{(N)}(r)=
\frac{-V_0}{1+\exp\left[\left(r-R_0-R_T \frac{\beta_2}{\sqrt{4\pi N}}
Q_{20}\right)/a\right]}. 
\label{vall}
\end{equation}
The matrix elements of Eq.~(\ref{vall}) can be evaluated most easily by 
introducing the interaction representation which diagonalizes the 
quadrupole operator $Q_{20}$ \cite{BBK94b}. 
Since phonon states $|n,I,M\rangle$ with $M\neq 0$ do not couple to the 
ground state in the no-Coriolis approximation, we do not need to consider 
the terms $d^{\dagger}_m d_m$ with $m\neq 0$ in the quadrupole operator, 
Eq.~(\ref{quad}).
Accordingly, we diagonalize the operator 
\begin{equation}
Q_{20}=s^{\dagger}d_0 + sd_0^{\dagger} + c_0 d_0^{\dagger}d_0,
\label{q20}
\end{equation}
where $c_0=-\sqrt{2/7}\chi_2$, in the 
interaction basis as
\begin{equation}
Q_{20}=e_+ B_+^{\dagger}B_+ + e_- B_-^{\dagger}B_-, 
\end{equation}
where $B_{\pm}^{\dagger}$ ($B_{\pm}$) are creation (annihilation) 
operators of the eigenbosons and $e_{\pm}$ are their 
eigenvalues.  After working out the commutators 
\begin{equation}
\left[Q_{20}, B_{\pm}^{\dagger}\right] = e_{\pm} B_{\pm}^{\dagger}, 
\end{equation}
we find that they are given by 
\begin{eqnarray}
B_+^{\dagger} &=& y_0 s^{\dagger} + y_2 d^{\dagger}_0, \\
B_-^{\dagger} &=& -y_2 s^{\dagger} + y_0 d^{\dagger}_0, \\
e_{\pm} &=& \frac{c_0 \pm \sqrt{4 + c_0^2}}{2},
\end{eqnarray}
with $y_0$ and $y_2$ are defined as 
\begin{eqnarray}
y_0 &=& \frac{1}{\sqrt{1+e_+^2}} = 
\sqrt{\frac{\sqrt{4+c_0^2}-c_0}{2\sqrt{4+c_0^2}}}, \\
y_2 &=& \frac{e_+}{\sqrt{1+e_+^2}} = 
\sqrt{\frac{\sqrt{4+c_0^2}+c_0}{2\sqrt{4+c_0^2}}}. 
\end{eqnarray}
The eigenvectors of the quadrupole operator are then given by 
\begin{equation}
|n_+,n_-> = \frac{1}{\sqrt{n_+!n_-!}}
(B_+^{\dagger})^{n_+} (B_-^{\dagger})^{n_-}|0>,
\end{equation}
and the matrix elements of the nuclear coupling Hamiltonian 
between $n$- and $m$- phonon channels read
\begin{eqnarray}
V_{nm}^{(N)}(r)
=&& \sum^N_{k=0} 
\frac{-V_0}{1+\exp\left[\left(r-R_0-R_T \frac{\beta_2}{\sqrt{4\pi N}}
(ke_+ + (N-k)e_-)\right)/a\right]} \nonumber \\
&& \times \sum_{i,j=0}^{k}(-)^{n+m+i+j}
\left(\begin{array}{c} k\\ i \end{array}\right)
\left(\begin{array}{c} N-k\\ N-n-i \end{array}\right)
\left(\begin{array}{c} k\\ j \end{array}\right)
\left(\begin{array}{c} N-k\\ N-m-j \end{array}\right) \nonumber \\
&& \times \frac{\sqrt{(N-n)!n!(N-m)!m!}}{k!(N-k)!}
y_0^{2i+2j-2k+m+n} y_2^{2k+2N-m-n-2i-2j}.
\label{vnm}
\end{eqnarray}
In evaluating Eq.~(\ref{vnm}), we used the fact that the sum of the 
number of each eigenfunction bosons, $n_+ + n_-$, must be equal to the total 
boson number $N$. 

\begin{center}
{\bf D. Coupling to octupole mode}
\end{center}

In spherical nuclei, the low-lying octupole vibrational state 
has a large collectivity and their excitations play an important 
role in subbarrier fusion reactions. Therefore, in order to apply 
the models which we discussed above to realistic systems, it is 
necessary to extend them so that they include the octupole mode as well. 
The coupling Hamiltonian in the harmonic limit can be 
extended in a straightforward way and reads 
\begin{equation}
V^{(N)}(r,\alpha)=
\frac{-V_0}{1+\exp\left[\left(r-R_0
-\sqrt{\frac{5}{4\pi}}R_T \alpha_{20}
-\sqrt{\frac{7}{4\pi}}R_T \alpha_{30}
\right)/a \right]},
\end{equation}
and 
\begin{equation}
V^{(C)}(r,\alpha)=\frac{3}{5}Z_PZ_Te^2\frac{R_T^2}{r^3}
\sqrt{\frac{5}{4\pi}}\alpha_{20} 
+ \frac{3}{7}Z_PZ_Te^2\frac{R_T^3}{r^4}
\sqrt{\frac{7}{4\pi}}\alpha_{30}, 
\end{equation}
for the nuclear and the Coulomb couplings, respectively. 
Here $\alpha_{20}$ and $\alpha_{30}$ are the surface coordinates 
for the quadrupole and the octupole 
vibrations, respectively. 
In order to take into account the octupole mode of 
excitation in the anharmonic oscillator coupling, we use 
the vibrational limit of the $sdf$- IBM \cite{IA87,BBWS88}. 
This model has been successfully used to describe negative parity 
states in rotational \cite{BBWS88} as well as 
vibrational \cite{BBD89,PBM90,JBJ93} nuclei. 
As a natural extension, we assume the following coupling 
Hamiltonians based on this model for the nuclear and the Coulomb 
couplings. 
\begin{eqnarray}
V^{(N)}(r,\xi)&=&
\frac{-V_0}{1+\exp\left[\left(r-R_0
-R_T \frac{\beta_2}{\sqrt{4\pi N}}Q_{20}
-R_T \frac{\beta_3}{\sqrt{4\pi N}}Q_{30}
\right)/a \right]}, \\
V^{(C)}(r,\xi)&=&\frac{3}{5}Z_PZ_Te^2\frac{R_T^2}{r^3}
\frac{\beta_2}{\sqrt{4\pi N}}Q_{20}
+ \frac{3}{7}Z_PZ_Te^2\frac{R_T^3}{r^4}
\frac{\beta_3}{\sqrt{4\pi N}}Q_{30}. 
\end{eqnarray}
Here, $\beta_3$ is the octupole deformation parameter, and 
we take the quadrupole and the octupole operators 
in the $sdf$- IBM as 
\begin{eqnarray}
&&\hat{Q}_2=s^{\dagger}\tilde{d} + sd^{\dagger} +
\chi_2(d^{\dagger}\tilde{d})^{(2)}
+ \chi_{2f}(f^{\dagger}\tilde{f})^{(2)}, 
\label{q2f} \\
&&\hat{Q}_3=sf^{\dagger} +
\chi_3(\tilde{d}f^{\dagger})^{(3)} + h.c.,
\label{q3f}
\end{eqnarray}
respectively, where $\tilde{f}_{\mu}$ is defined as 
$(-)^{3+\mu}f_{-\mu}$. 
We will apply these models in Sec. IV to analyze the 
$^{16}$O + $^{148}$Sm fusion reaction. 

\begin{center}
{\bf III. ANHARMONIC EFFECTS ON FUSION BARRIER DISTRIBUTIONS}
\end{center}

In order to discuss the effects of anharmonicity 
on subbarrier fusion reactions, 
we perform in this section a series of model calculations without 
particular regard for the physical values of the parameters. 
We consider a fictitious nucleus whose excitation energies are 
given by Eq. (\ref{eu5}) with $\epsilon=800$, $\alpha=10$, $\beta=0$ 
and $\gamma=25$, all in keV. The excitation energy of the first 2$^+$ 
state of this nucleus is 1 MeV and the double phonon states are split 
by 0.5 MeV. We take the total boson number $N$ to be 4 and 
thus take into account coupling of up to four phonon states. 
The other parameters used are $\chi_2=-3$ (see Eq.~(\ref{quad})), 
the quadrupole deformation parameter $\beta_2=0.25$ and the 
target radius $R_T=6.35$ fm. 
Finally, the Christensen-Winther parameterization of 
the Woods-Saxon potential \cite{CW76} for the $^{16}$O + $^{148}$Sm 
system is used for the ion-ion potential $V_N$.  

As we discussed in the previous section, introduction of the 
multi-phonon channels significantly reduces the dimension of the 
coupled-channels equations.  To see whether this approximation is good, 
a calculation is first made by taking fully into account the effects of 
the splittings in the phonon spectrum.  In order to keep the calculation 
simple, we make here the linear coupling approximation.  In Fig.~2, we 
show the excitation function of the fusion cross section (the upper 
panel) and the barrier distribution (the lower panel) that follow from 
using the U(5) limit as described above (solid line).  Barrier 
distributions are obtained using a point-difference formula \cite{LDH95} 
with $\Delta E$ =2 MeV in the laboratory frame.  To see the effect of 
the splittings of the multi-phonon states, we have repeated the same 
calculation by assuming that the $n$-phonon multiplets are degenerate at 
the excitation energy of the first $(2n)^+$ state.  We found that this 
prescription leads to no visible change in the fusion barrier 
distribution compared to the actual case with splitt energy levels.  
In the following, we shall therefore assume a degenerate multi-phonon 
spectrum and use a single multi-phonon channel representing all the 
states in a multiplet.

In Fig.~2, we also compare the results for the anharmonic vibrator with 
those in the harmonic limit (the dotted line).  One can 
observe that the anharmonic effects in the vibrational excitations lead 
to a significant change in the barrier distribution, though the 
excitation function of the fusion cross section 
itself is modified only marginally. 

\begin{center}
{\bf A. Anharmonicities in excitation energies}
\end{center}

In all the following calculations, we take into account the couplings to all 
orders.  We first discuss the effects of the deviation of the energy 
spectrum from the harmonic limit.  In order to isolate these effects, we 
truncate the phonon spectrum at the double-phonon states in this 
subsection.  Fig.~3 shows the dependence of the fusion cross section 
(the upper panel) and the fusion barrier distribution (the lower panel) 
on the parameter $\delta$ in Eq.~(\ref{v2p}).  The cross section and the 
barrier distribution are calculated for three different values of 
$\delta$.  The solid line is the result when $\delta=0$, while the 
dotted and the dashed lines are obtained by setting $\delta$ to be 0.5 
and $-$0.5 MeV, respectively.  Despite the fact that an unrealistically 
large value is used for $\delta$ so as to maximize its effect, the 
barrier distribution changes only slightly for different choices of 
$\delta$.  This indicates that the main effects of anharmonicity on 
fusion barrier distributions come from the deviation of the transition 
probabilities from the harmonic limit, including the reorientation 
effect, rather than anharmonicities in the level energies.  
Note that this 
observation does not necessarily mean that fusion reactions are not 
sensitive to the excitation energy of the phonon states.  To demonstrate 
this, we show in Fig.~3 also the result when both $\hbar\omega$ and 
$\delta$ is set to be zero (the dot-dashed line).  One can see sizable 
effects of the finite excitation energy of the phonon states on the 
fusion barrier distribution as well as fusion cross section.  We thus 
conclude that, though the fusion cross section and the barrier 
distribution are sensitive to the energy of the single-phonon state, 
they do not depend so much on the excitation energies of the 
multi-phonon states once the phonon quanta $\hbar\omega$ is fixed.

\begin{center}
{\bf B. Reorientation effects}
\end{center}

One of the pronounced features of an anharmonic vibrator is that the 
excited states have non-zero quadrupole moments \cite{TU66,BM75}. 
In the IBM, the E2 operator is defined as $T$(E2$)=e_B \hat Q_2$, where 
the quadrupole operator was introduced in Eqs.~(\ref{quad}) and 
(\ref{q2f}) for the $sd$- and $sdf$-IBM, respectively, 
$e_B$ being is the effective charge. 
In the U(5) limit, the E2 effective charge $e_B$ 
is related to the E2 transition probability by 
\begin{equation}
B(E2;0_1^+ \to 2^+_1) = 5e_B^2 N. 
\label{be2}
\end{equation}
Using the definition of the static quadrupole moment of a state with spin 
$I$ 
\begin{equation}
Q(I)=\sqrt{\frac{16\pi}{5}} <II | T(E2) | II>,
\end{equation}
we obtain for the quadrupole moment of the 2$^+$ state 
\begin{equation}
Q(2^+)=\sqrt{\frac{32\pi}{35}} e_B \chi_2.
\label{chi}
\end{equation}
The corresponding formula for the first 3$^-$ state reads
\begin{equation}
Q(3^-)=\sqrt{\frac{20\pi}{21}} e_B \chi_{2f}.
\label{chif}
\end{equation}
Fig.~4 shows the influence of the sign of the quadrupole moment of the 
first 2$^+$ state on the fusion cross section (the upper 
panel) and the fusion barrier distribution (the lower panel).  The solid 
line corresponds to the negative static 
quadrupole moment, while the dotted line is obtained by inverting the 
sign of the $\chi_2$ parameter in Eq.~(\ref{quad}).  As seen from 
Eq.~(\ref{chi}), this is equivalent to taking the opposite sign for the 
quadrupole moment of the excited state.  The dashed line is the result 
when $\chi_2=0$.  Fig.~4 demonstrates that the fusion cross section, 
and especially fusion barrier distribution, strongly depend on the sign 
of the $\chi_2$ parameter, hence on that of the static quadrupole moment.  
This fact will be used in the next section to determine the quadrupole 
moments of the first 2$^+$ and 3$^-$ states in $^{148}$Sm from the 
experimental fusion barrier distribution for the $^{16}$O + $^{148}$Sm 
reaction.

\begin{center}
{\bf C. Finite $N$ effects}
\end{center}

The other important effect of the deviation of the electric transition 
rates from the harmonic limit is caused by the finiteness of the boson 
number.  Eq.~(\ref{v2p}) and Fig.~1 indicate that the anharmonic 
effects weaken the coupling between the one- and two-phonon states by 
the factor $\sqrt{1-1/N}$ compared with that in the harmonic limit.  For 
small values of the boson number $N$, this factor is significantly 
smaller than one and large anharmonic effects on fusion reactions are 
expected.  In order to demonstrate the finite boson number effects, we 
show in Fig.~4 also the results in the harmonic limit by the dot-dashed 
line.  One can see a significant difference between this result and that 
for the zero quadrupole moment (the dashed line), 
indicating the importance of the finite boson number effect.

Finite boson number effects can also be studied in another, perhaps more 
instructive way.  Since the couplings to the multi-phonon states are 
weaker than those in the harmonic limit if the anharmonic effects are 
present, one expects that fusion excitation function converges more 
rapidly compared with the case in the absence of the anharmonic effects.  
Fig.~5 shows how the fusion cross sections (the upper panel) and the 
fusion barrier distributions (the lower panel) converge with 
the phonon number for fixed total boson number $N=4$.  Although the 
fusion cross section and the fusion barrier distribution obtained by 
including only the single-phonon excitations are quite different from 
those in the double-phonon couplings, the difference between the two- 
and the three-phonon couplings are small.  We observe that the fusion 
barrier distribution almost converges at the three-phonon level.  The 
corresponding calculations in the harmonic limit are performed for 
comparison.  These results are shown in Fig.~6.  Contrary to the 
results for the anharmonic vibrator, the barrier distribution obtained 
by taking into account up to the four-phonon excitations is still 
significantly different from that obtained by including up to the 
three-phonon excitations.  
We thus conclude that the finite boson number 
significantly influences the role of higher phonon states 
in determining the fusion barrier distribution.

\begin{center}
{\bf IV. $^{16}$O + $^{148}$Sm REACTION: COMPARISON WITH \\ 
EXPERIMENTAL DATA}
\end{center}

We now apply the formalism described in Sec. II 
to a realistic system.  We analyze in 
particular the $^{16}$O + $^{148}$Sm reaction, whose excitation function 
of the fusion cross section was recently measured with high accuracy 
\cite{LDH95}.  
We take both the quadrupole and the octupole vibrational excitations 
in $^{148}$Sm into account, 
while the excitations of $^{16}$O are not explicitly 
included in the coupled-channels calculations.  
The latter have been shown to 
lead only to the shift of the fusion barrier distribution in energies 
without significantly altering its shape \cite{HTD97b} and hence 
can be incorporated in the choice of the bare potential. 

In order to qualitatively understand the effects of the channel 
couplings on this reaction, calculations are first performed by 
assuming the harmonic limit.  The quadrupole and the octupole 
deformation parameters are estimated from the experimental electric 
transition probabilities between the ground and the one-phonon 
states \cite{P90} to be $\beta_2=0.182$ and $\beta_3=0.236$, 
respectively, by assuming the target radius of $R_T=1.06 A^{1/3}$ fm.  
The excitation energies of the phonon states in $^{148}$Sm are 0.55 and 
1.16 MeV, for the first 2$^{+}$ and 3$^{-}$ states, respectively. The 
parameters for the bare ion-ion potential are obtained by fitting the 
experimental fusion cross sections.  It has been pointed out that the 
effects of channel couplings play an important role in determining the 
bare potential for the $^{16}$O + $^{144}$Sm reaction \cite{EB96}.  
Accordingly, we fit the experimental data by assuming the three 
phonon couplings in the harmonic limit (see the following discussion).  
We use the experimental data between 200 mb and 400 mb to determine the 
bare potential.  We choose this range because, 
more details of the channel couplings would be important
in the lower energy region, 
while some other effects, e.g.  the angular momentum 
truncation or the dissipation mechanism might play some role \cite{VAS81} 
in the higher energy region.  
The best fit parameters which we obtain are $V_0=155.1$ 
MeV, $R_0 = 0.95 (A_P^{1/3} + A_T ^{1/3})$ fm, and $a=1.05$ fm, for the 
depth, the radius, and the diffuseness parameters of the 
Woods-Saxon potential, respectively.  Note that the experimental data at 
high energy region still require a large surface diffuseness parameter 
\cite{LDH95} even after including the effects of the channel couplings.  
The origin of this is not fully understood.

The results for the fusion barrier distribution for the $^{16}$O + 
$^{148}$Sm reaction are shown in Fig.~7.  The panels differ from each 
other by the number of the quadrupole phonons coupled.  The experimental 
data are taken from ref. \cite{LDH95}.  The point difference formula 
with $\Delta E$ = 2 MeV in the laboratory frame is used to obtain the 
fusion barrier distribution.  Calculations, which assume that the 
octupole vibration in $^{148}$Sm has only a single-phonon state, are 
indicated by the dotted lines in Fig.~7.  In the three phonon 
calculation, for example, we included the 2$^{+}$, 3$^-$, 2$^+ 
\bigotimes 2^+$, 2$^+ \bigotimes 3^-$, 2$^+ \bigotimes 2^+ \bigotimes 
2^+$, and 2$^+ \bigotimes 2^+ \bigotimes 3^-$ states in $^{148}$Sm in 
the harmonic limit.  The coupling scheme in the other panels is defined 
in the same way.  All the dotted lines which are obtained by assuming 
the single-octupole-phonon excitations in $^{148}$Sm do not account for 
the experimental fusion barrier distribution.  However, double-phonon 
excitations have been found in the neighbouring nuclei, i.e.  $^{144}$Sm 
\cite{GVB90,GJB93,WRZB96,MHK93}, $^{146}$Sm \cite{BBB95}, and $^{147}$Sm 
\cite{UBJ96}.  We therefore repeat the same calculations by assuming the 
double-octupole excitations in $^{148}$Sm (the solid lines).  One can 
now observe that the experimental data can be explained reasonably well 
by the three-phonon calculations.  The fit is not as good if the four 
phonon quadrupole excitations are included in the coupled channels 
analysis. These calculations in the harmonic limit thus suggest that 
there are strong couplings up to the three quadrupole phonon states in 
$^{148}$Sm and the coupling between the three and four phonon states are 
not as strong as that expected from the harmonic oscillator coupling.  
They also suggest that there might be double-octupole phonon 
excitations in $^{148}$Sm.

Calculations which take into account the anharmonicities of the 
vibrational excitations are performed next.  
Following the results of the analysis in the harmonic oscillator 
limit, we include the octupole 
excitations up to the double phonon levels 
and take the total boson 
number $N=4$.  The latter is consistent with the 
existence of the $Z=64$ subshell closure \cite{CW88}.  
The standard 
prescription for the boson number (i.e.~counting pairs of nucleons above 
or below the nearest shell closure) would give $N=8$ for $^{148}$Sm.  
However, it is well known that due to the $Z=64$ subshell closure, the 
effective boson number is much smaller.  The suggested effective 
number in the literature for the proton bosons varies 
between $N_{\pi}=1$ and 3 \cite{WC87,S83}, leading 
to the total boson number to be 
between 3 and 5 for $^{148}$Sm.  
There are three other parameters, i.e.  
$\chi_2$, $\chi_{2f}$, and $\chi_3$ parameters in the transition 
operators Eqs.~(\ref{q2f}) and (\ref{q3f}), which need to be determined.  
Two of them, $\chi_2$ and $\chi_{2f}$, are directly related to the 
quadrupole moment of the phonon states, and the other the coupling 
between the quadrupole and the octupole modes.  Though there exist 
experimental data on the quadrupole moment of the first 2$^+$ state of 
$^{148}$Sm \cite{P90}, the experimental uncertainty is still large.
 There are no other experimental data which can be used to fix the 
$\chi_{2f}$ and $\chi_3$ parameters.  Therefore, a $\chi^2$ fit to the 
experimental fusion cross sections is carried to determine all the three 
parameters.  In the fitting procedure, all the experimental data below 
200 mb are used except for the one at E$_{lab}$= 69.36 MeV, 
which lies outside the systematics.  
The best fit values are $\chi_2 = 
-3.12 \pm 0.77, \chi_{2f} = 4.63 \pm 0.43$, and $\chi_3 = -1.99 \pm 
0.26$.  
The solid line in Fig.~8 shows the fusion cross section (the 
upper panel) and the fusion barrier distribution (the lower panel) thus 
obtained.  Following the discussion in the previous section, harmonic 
spectra for the excitation energies of the phonon states are assumed.  
One observes that the shape of the fusion barrier distribution 
obtained by including 
up to four phonon states with anharmonicities 
is very 
similar to that for the three phonon couplings in the harmonic limit 
(see Fig.~7) and that they agree quite well with the experimental data.
This situation is similar to that observed in ref. \cite{HTK97}, where it 
has been shown that the shape of the fusion barrier distribution for the 
$^{16}$O + $^{144}$Sm reaction obtained by including up to single 
phonon states in the harmonic limit is very similar to that for the 
double phonon couplings with anharmonicities. 

In the last section, we discussed two main effects of the anharmonicities 
in the electric transitions, i.e. the reorientation  
and the finite $N$ effects on subbarrier fusion reactions. In order 
to see each effect separately, we repeat the same calculation by 
setting both the $\chi_2$ and $\chi_{2f}$ parameters to be zero. 
As was discussed in the previous section, this prescription is 
equivalent to taking only the finite $N$ effects into 
account. The result is shown by the dashed line in Fig.~8. 
The figure also shows for comparison the result 
obtained by assuming that there are four quadrupole 
phonon states in the harmonic oscillator limit in $^{148}$Sm 
(the dotted line). 
The difference between the dotted and the dashed lines, and also 
between the dashed and the solid lines, is significant. 
This suggests that both the reorientation and the finite $N$ 
effects play an important role in the fusion reaction. 

From the $\chi_2$ and $\chi_{2f}$ parameters which are obtained 
by the $\chi^2$ fit to the experimental fusion cross sections, 
we estimate the quadrupole moments of the first 2$^+$ and 3$^-$ 
states of $^{148}$Sm. The E2 effective charge $e_B$ is estimated
to be 0.19 $eb$ from the experimental $B(E2)$ value (see Eq.~(\ref{be2})). 
Using Eqs.~(\ref{chi}) and (\ref{chif}), 
they are estimated to be 
$Q(2^+) = -1.00 \pm 0.25~b$ and 
$Q(3^-) = +1.52 \pm 0.14~b$, respectively. 
The quadrupole moment of the first 2$^+$ state thus obtained 
is very close to that measured from the Coulomb 
excitation technique, i.e.  $-0.97 \pm 0.27 b$ \cite{P90}.  
There is no experimental data to compare our result 
for the quadrupole moment of the first $3^-$ state. 
We can, however, test the 
consistency of the fit by taking its opposite sign in the 
coupled-channels calculations.  Fig.~9 shows the sensitivity of the 
fusion cross section and the fusion barrier distribution to the sign of 
the quadrupole moment of the first 3$^-$ state.  The solid line 
corresponds to the optimal choice for the sign of the first 3$^-$ state, 
while the dotted line is obtained by inverting it.  The 
use of the incorrect sign of the quadrupole moment destroys the good fit 
to the experimental data.  Our calculations thus strongly suggest that 
the heavy-ion fusion reactions at low energies
can provide an alternative method to determine the 
sign as well as the magnitude of the quadrupole moments 
of the low-lying excited states in spherical nuclei.

\begin{center}
{\bf V. SUMMARY AND CONCLUDING REMARKS}
\end{center}

We discussed the effects of anharmonic phonon excitations 
on heavy-ion fusion reactions at subbarrier energies and showed 
that they play an important role. 
We showed that the vibrational limit of the interacting boson 
model provides a useful framework to address these questions. 
There are mainly three effects of anharmonicities; the anharmonicity 
in the excitation energy spectra, the reorientation effects, 
and the finite boson number effects in the transition strength. 
We showed that the anharmonic effects in the energy spectra 
play only a minor role and the main effects 
come from the deviation of transition probabilities from the harmonic 
limit, i.e. both the reorientation and the finite $N$ effects. 
We found that the fusion barrier distribution strongly depends on 
the sign of the quadrupole moment of excited states. 
Using these properties, we have analyzed the 
$^{16}$O + $^{148}$Sm fusion reaction and discussed the anharmonic 
properties of the phonon excitations in $^{148}$Sm. 
It was found that 
the best fit to the experimental data requires a negative and a positive 
quadrupole moments for the first 2$^+$ and 3$^-$ states of $^{148}$Sm, 
respectively. 
Since the quadrupole moment of the first 3$^{-}$ state of $^{144}$Sm 
has been found to be negative \cite{HTK97}, it would be interesting 
to measure that of the nucleus in between, i.e. $^{146}$Sm. 

We have shown that calculations in the harmonic limit provide 
only qualitative results and the realistic situations are much 
more complex due to anharmonicities. 
Since harmonic calculations suggested that a fusion barrier 
distribution is quite sensitive to the number of phonon states excited 
during the fusion, it has been hoped that subbarrier fusion 
reactions offer an alternative method to identify the existence 
of higher phonon states. 
Actually, some analyses, aiming at identifying multi-phonon 
states, have been carried out. 
It is, however, apparent from our results that care has to be taken 
in such analyses. 
Our calculations show that fusion barrier distribution 
converges more rapidly with the number of phonons when the anharmonic 
effects are present 
compared with the case in their absence. 
Harmonic calculations may thus underestimate the maximum number of 
phonons in vibrational nuclei. 
It is vital to take into account anharmonic effects 
in order to properly identify 
the maximum number of phonons. 

\begin{center}
{\bf ACKNOWLEDGMENTS}
\end{center}

The authors thank J.R. Leigh, M. Dasgupta, and D.J. Hinde for
useful discussions and comments.  
The work of K.H. was supported by the Japan Society for the Promotion of
Science for Young Scientists.  
This work was also supported by the Grant-in-Aid for General                
Scientific Research,                                                        
Contract No.08640380, Monbusho International Scientific Research Program:   
Joint Research, Contract No. 09044051,                                      
from the Japanese Ministry of Education, Science and Culture.

\newpage

\begin{center}
{\bf Figure Captions}
\end{center}

\noindent
{\bf Fig. 1:}
Schematic representation of the coupling scheme in vibrational 
models. The upper part is for the anharmonic vibrator, while the 
lower part for the harmonic oscillator. See text for the notation. \\

\noindent 
{\bf Fig. 2:} 
Effects of anharmonicity of the quadrupole vibration on the 
excitation function of the fusion 
cross section (the upper panel) and fusion barrier distribution 
(the lower panel). 
The dotted line is the result in the harmonic limit, while the solid 
line takes the anharmonic effects of the vibrational 
excitations into account. The linear order coupling is assumed. \\

\noindent
{\bf Fig. 3:}
Dependence of the fusion cross section (the upper panel) and fusion barrier 
distribution (the lower panel) on the deviation 
of the energy spectrum from the harmonic limit. The solid, 
the dotted, and the dashed lines are obtained by setting the 
parameter $\delta$ in  Eq.~(\ref{v2p}) to be 
zero, 0.5, and $-$0.5 MeV, respectively. The dot-dashed line is the 
result in the degenerate limit in the anharmonic vibrator 
coupling. The phonon spectrum is truncated at the double phonon 
levels. The full order couplings are included. 
\\

\noindent
{\bf Fig. 4:}
Dependence of the fusion cross section (the upper panel) and 
fusion barrier distribution (the lower panel) on the 
sign of the quadrupole moment of the first 2$^+$ state. 
The solid and the dotted lines correspond to the cases 
for the negative and the positive static quadrupole moments, 
respectively. 
The dashed line was obtained by setting the quadrupole moment 
of the excited states to be zero. 
The dot-dashed line is the result of the corresponding 
calculations in the harmonic limit. \\

\noindent
{\bf Fig. 5:}
Convergence of the fusion 
cross section (the upper panel) and fusion 
barrier distribution (the lower panel) 
as functions of the 
number of phonon states included in the coupled-channels 
equations, which are indicated in the inset. 
The total boson number $N$ is fixed to be four in all 
cases. \\

\noindent
{\bf Fig. 6:}
The same as Fig. 5, but for 
the harmonic oscillator coupling. \\

\noindent
{\bf Fig. 7:}
Comparison of the experimental fusion barrier distribution with
the results of coupled-channels calculations in the harmonic limit 
for the $^{16}$O + $^{148}$Sm reaction.  
The dotted lines include the single octupole excitations, while 
the solid lines take the double octupole phonon couplings 
into account. 
The experimental data are taken from Ref. \cite{LDH95}.  \\

\noindent
{\bf Fig. 8:}
Comparison of the experimental fusion cross section
(the upper panel) and fusion barrier distribution (the lower panel) with
the coupled-channels calculations for the $^{16}$O + $^{148}$Sm reaction.  
The experimental data are taken from Ref. \cite{LDH95}.  
The anharmonic effects in the quadrupole and the octupole vibrational 
excitations in $^{148}$Sm are taken into account in the $sdf$- IBM.
The solid line is the result when the best fit parameters are used, 
while the dashed line 
is obtained by setting the quadrupole moments of all the excited 
states to be zero. 
The results including the four phonon couplings in the harmonic limit is 
denoted by the dotted line. \\

\noindent
{\bf Fig. 9:}
Dependence of the fusion cross section (the upper panel) and 
fusion barrier distribution (the lower panel) on the 
sign of the static quadrupole moment of the first 3$^-$ state 
for the $^{16}$O + $^{148}$Sm reaction.  
The solid and the dotted lines correspond to the cases 
where the quadrupole moment of the 
first 3$-$ state in $^{148}$Sm is positive and negative, respectively. 

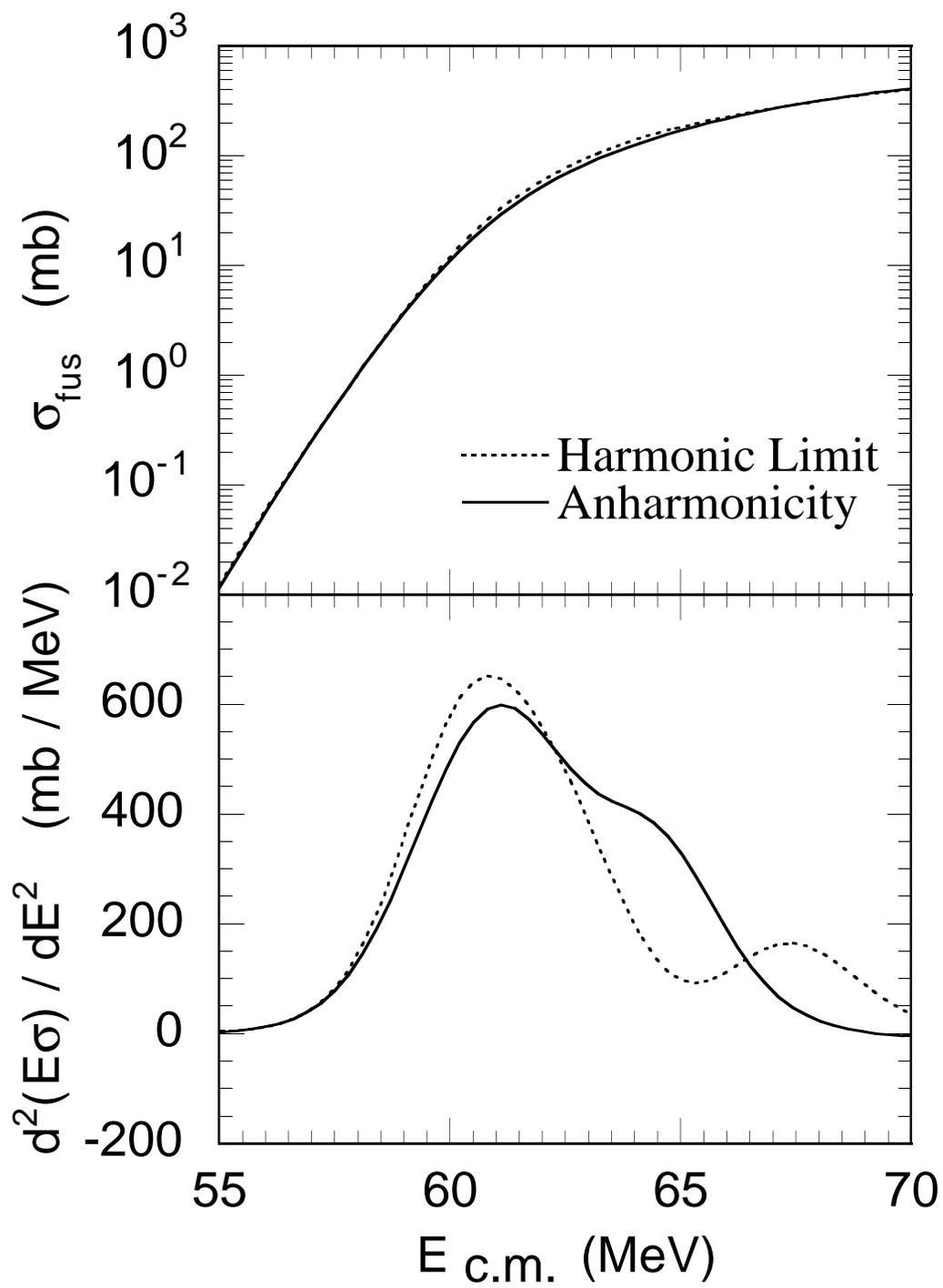
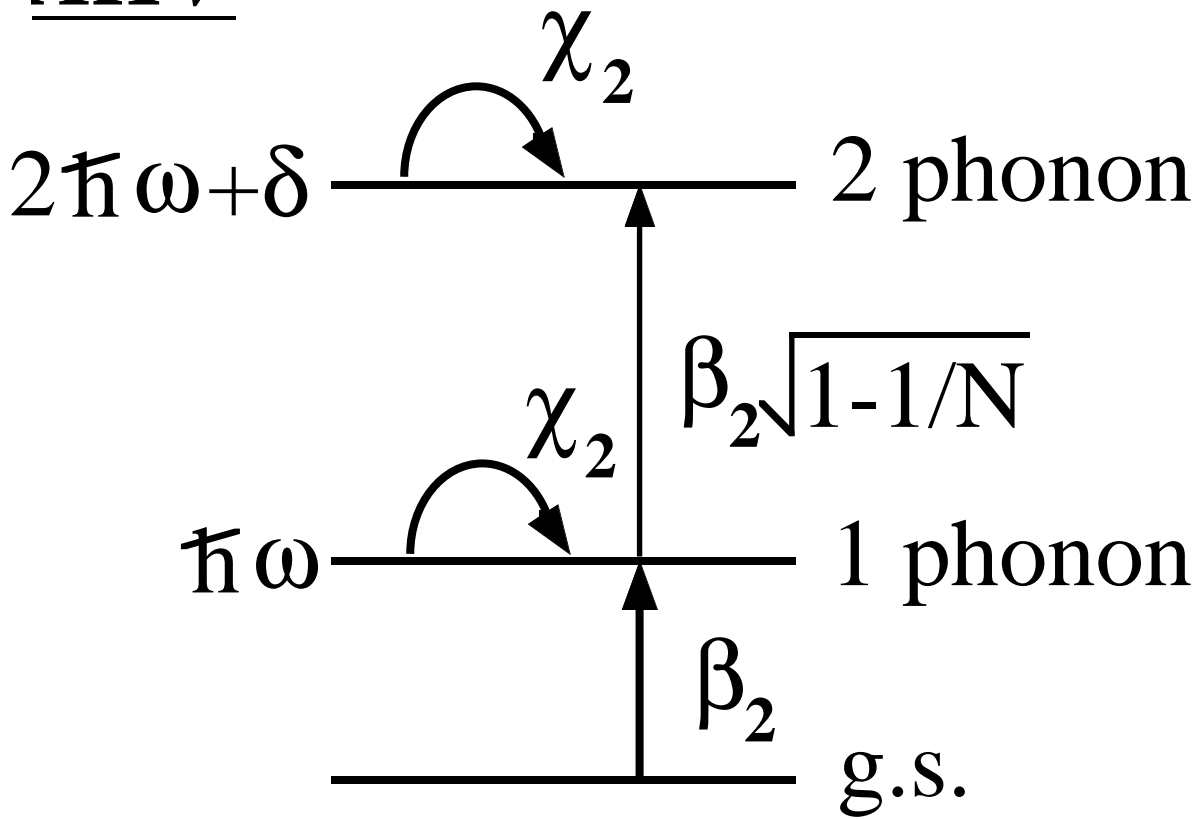
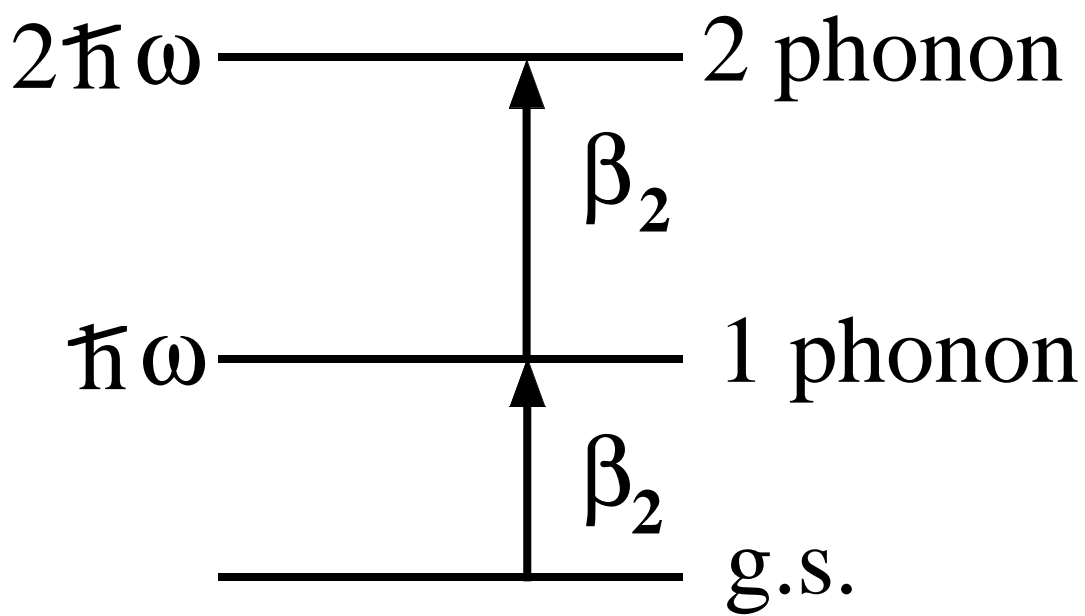
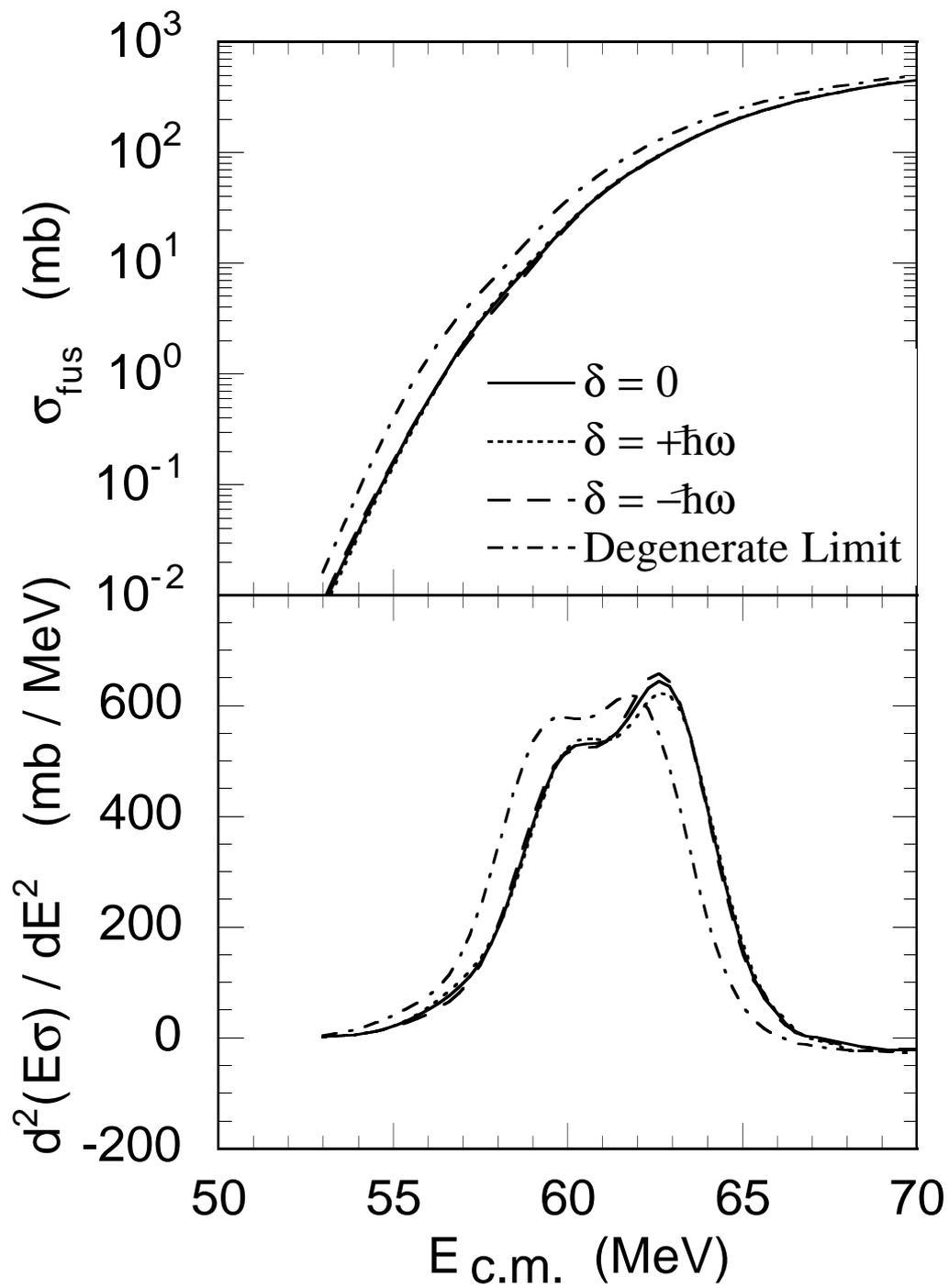
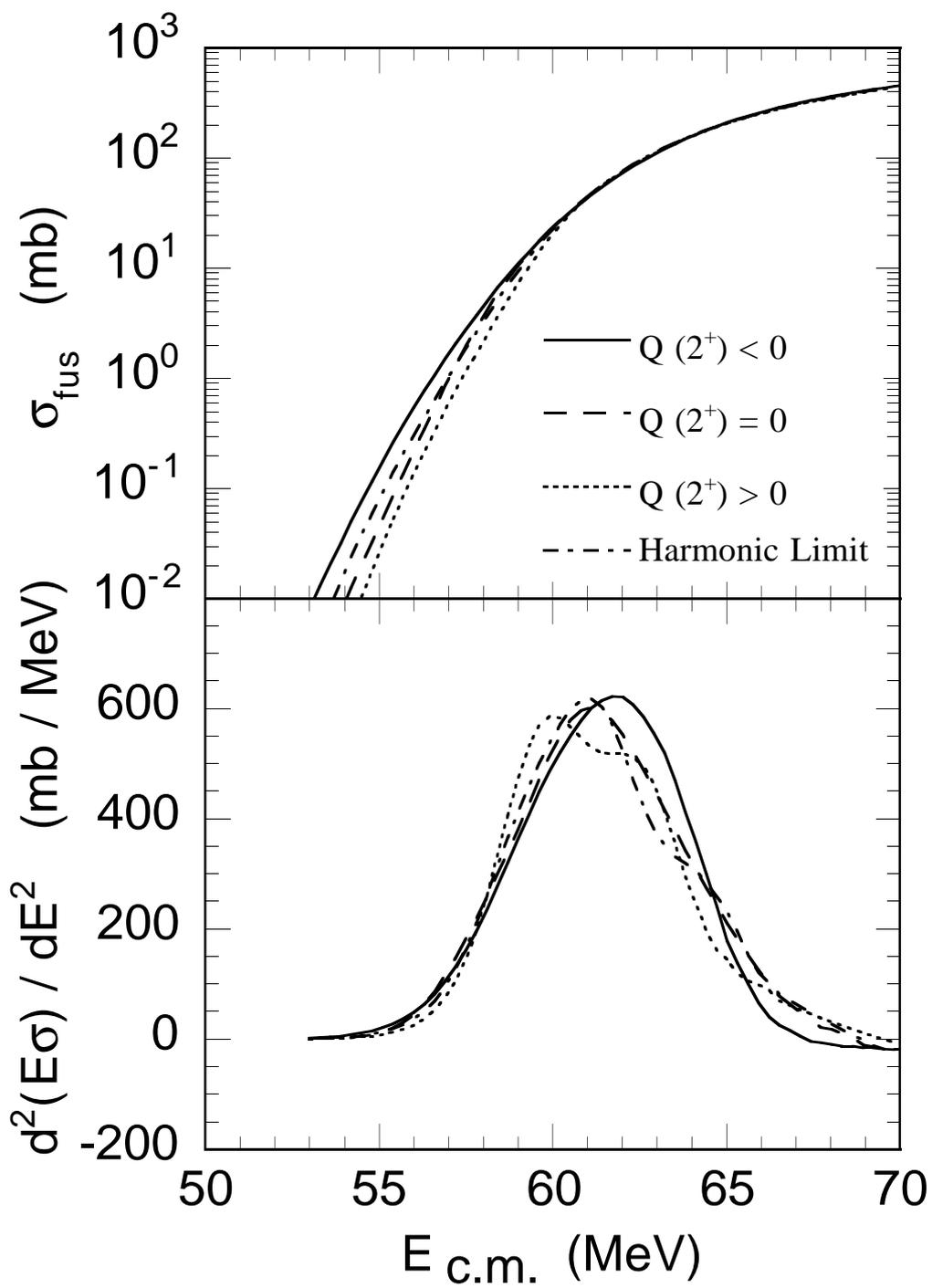
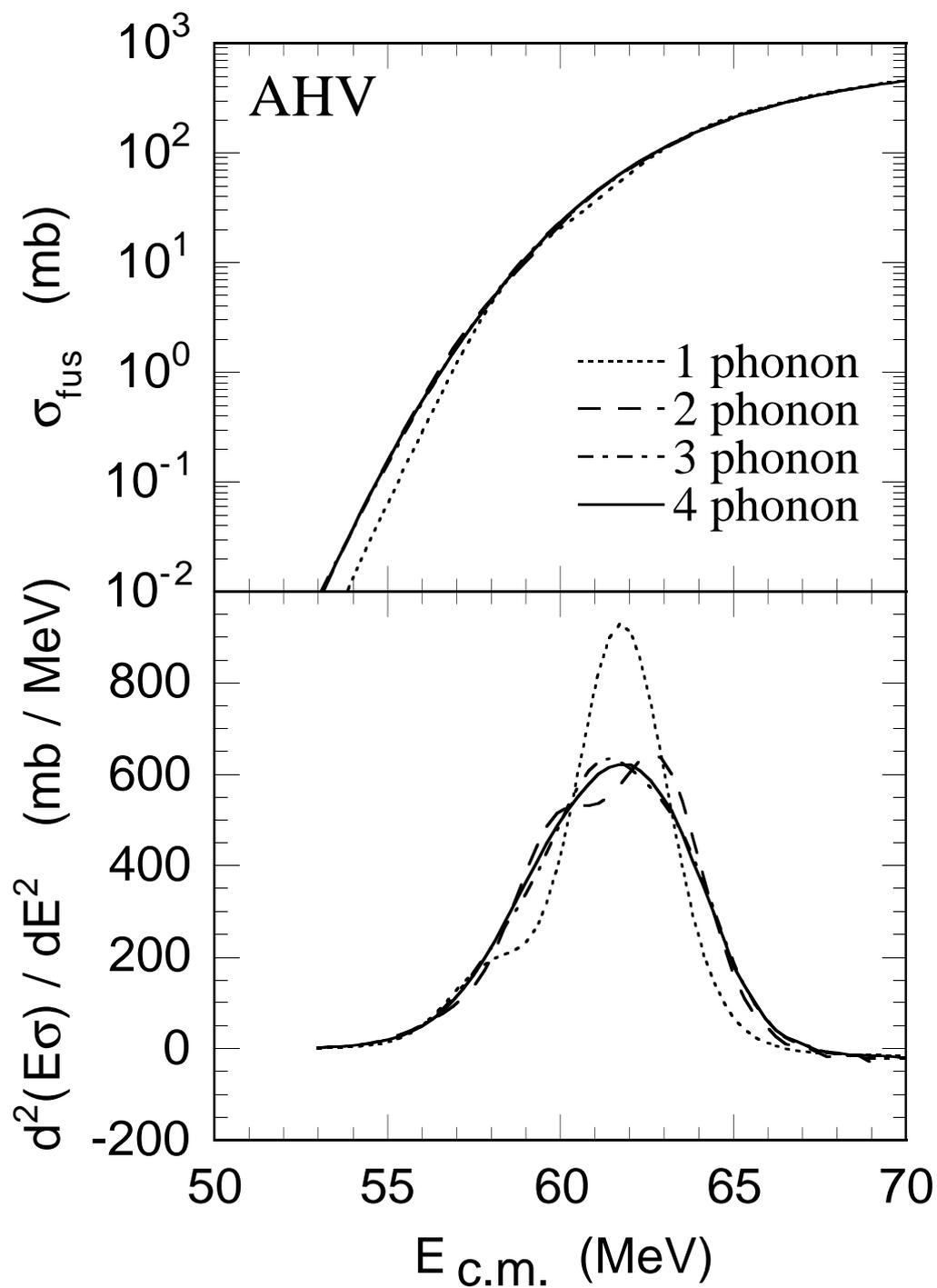
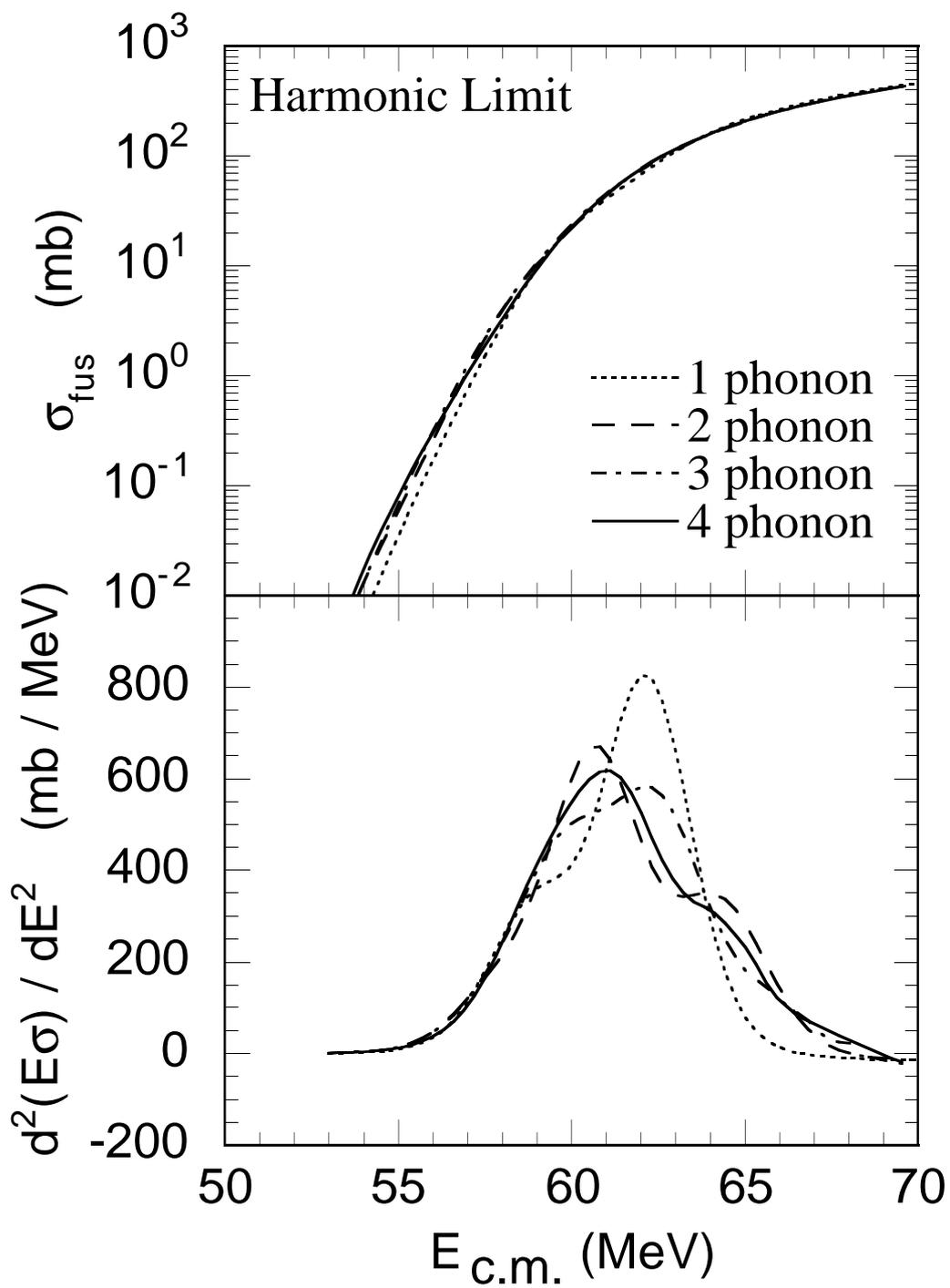
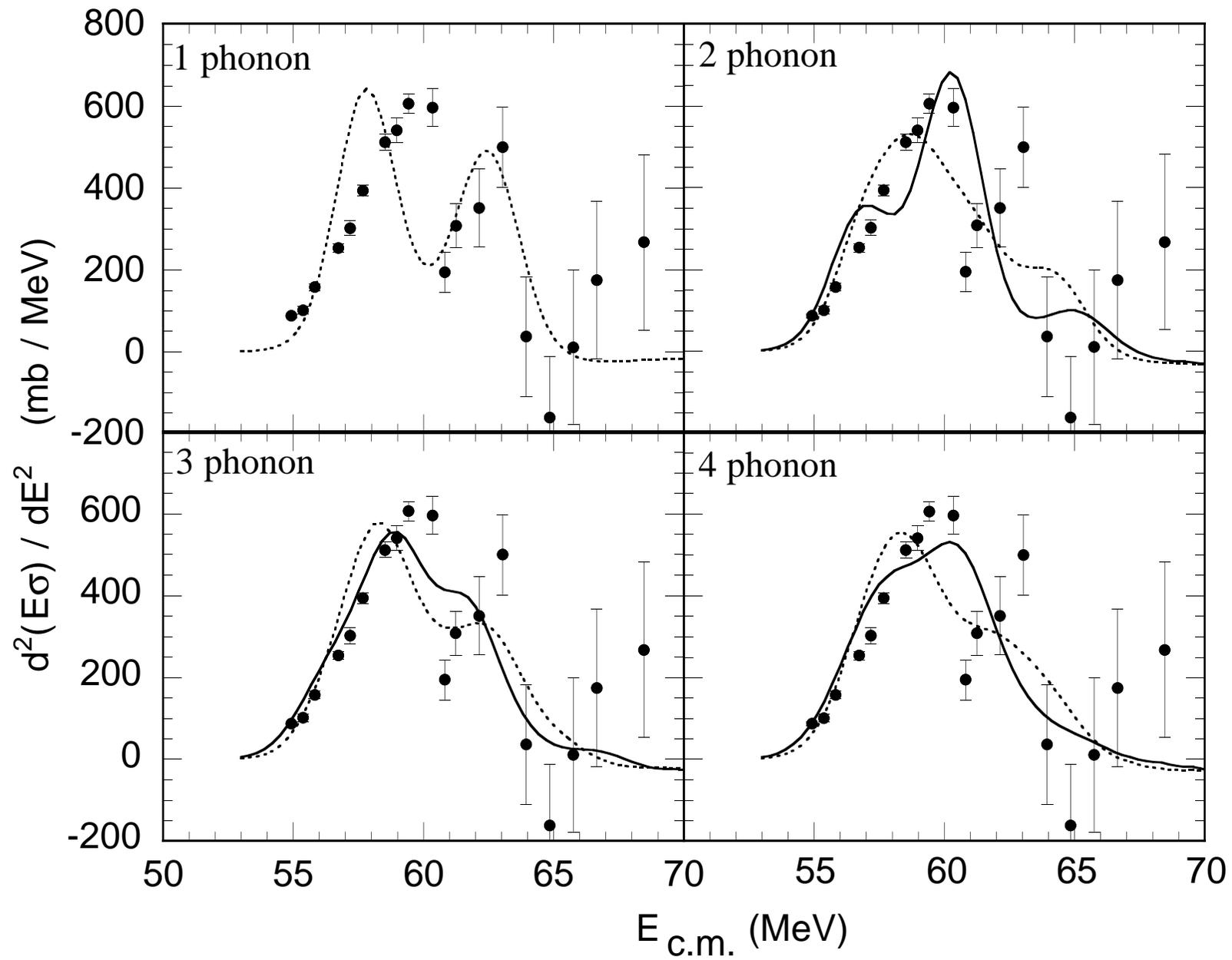
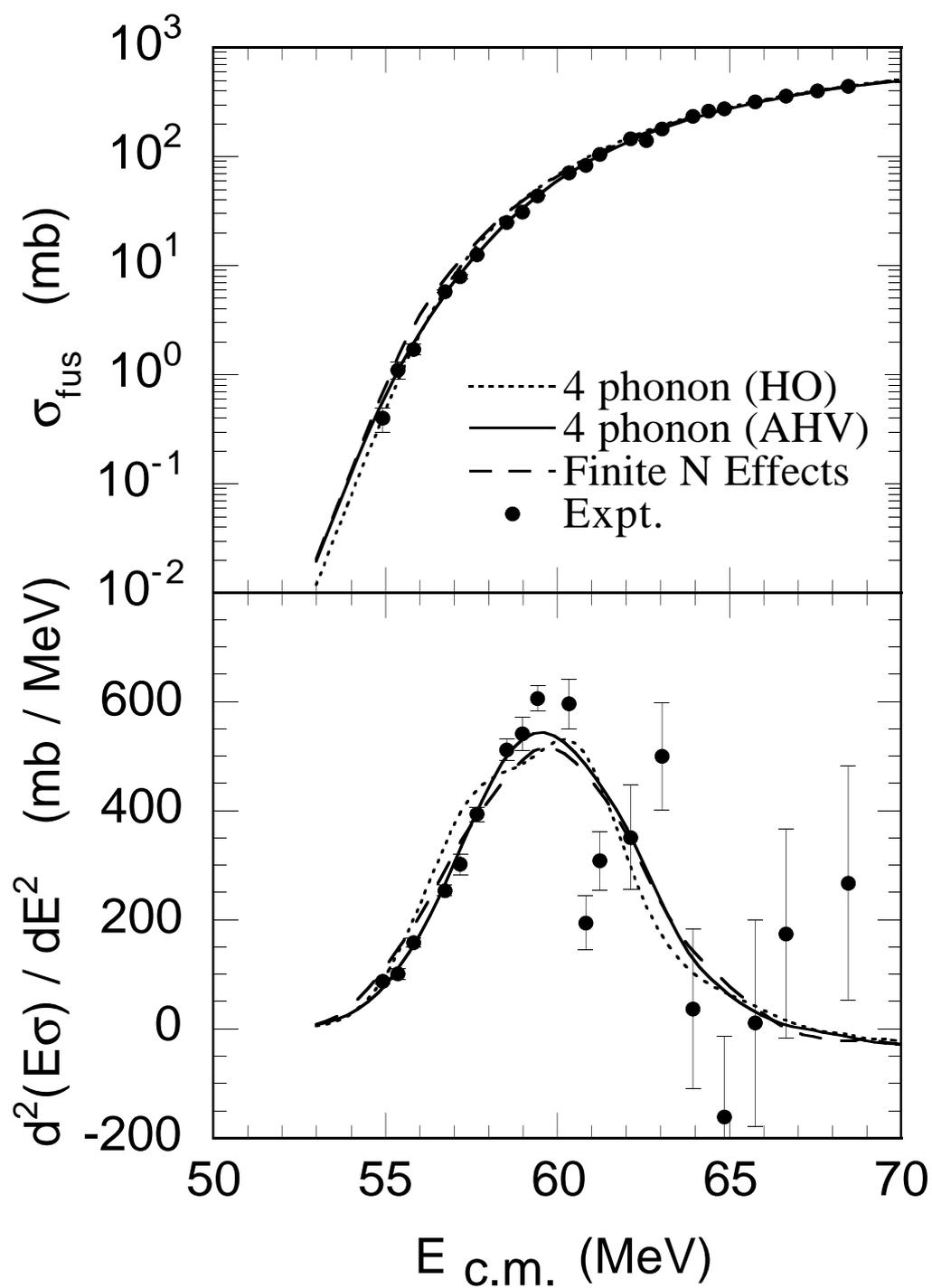
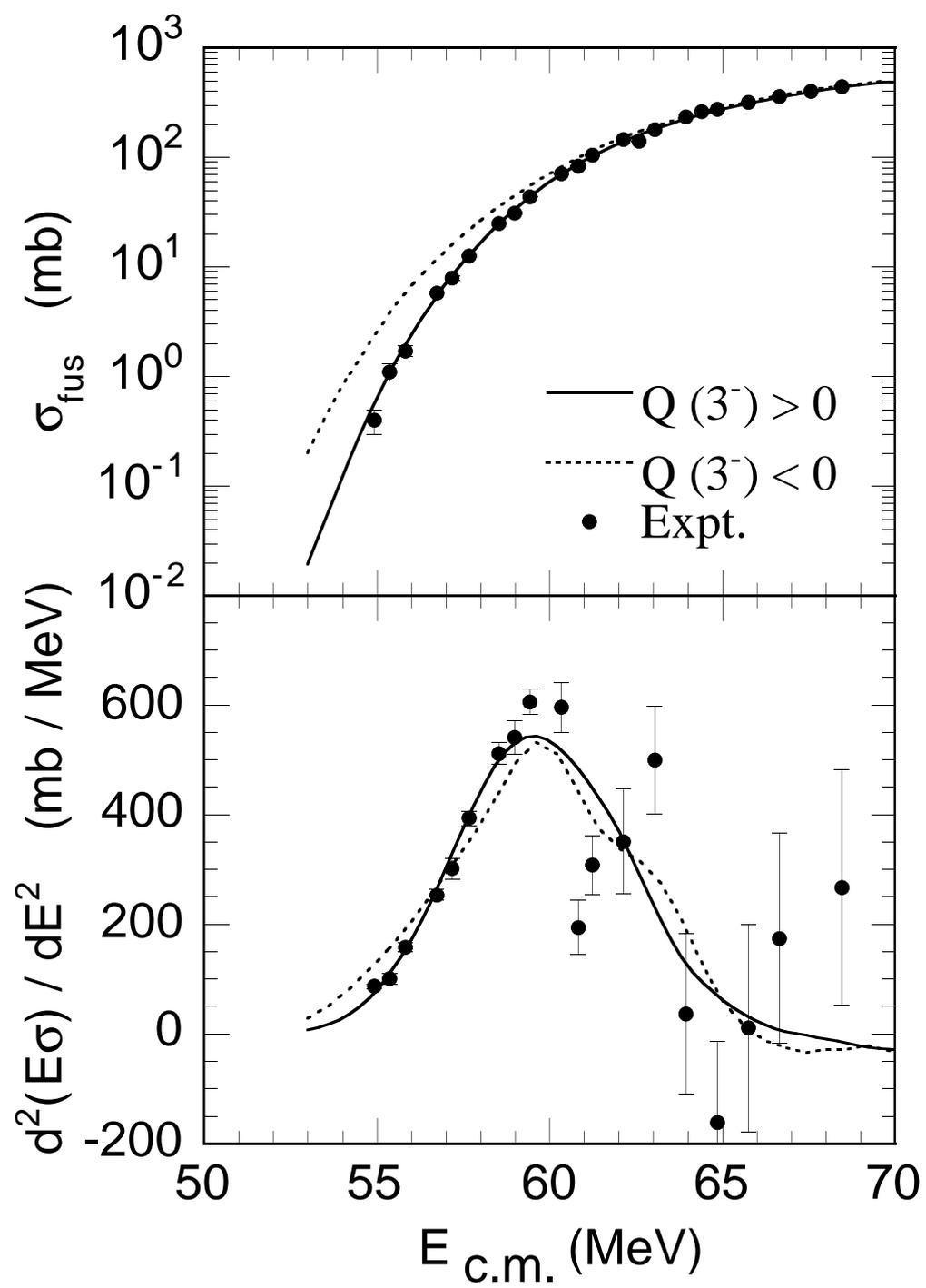
\end{document}